# Anisotropic Frictional Response of Texture Induced Strained Graphene

Andrea Mescola[1#], Guido Paolicelli[1#], Roberto Guarino[2], James G. McHugh[3], Sean P. Ogilvie[4], Alberto Rota[1,5], Enrico Gnecco[6], Erica Iacob[7], Sergio Valeri[1,5], Nicola M. Pugno[8,9], Venkata Gadhamshetty[10], Muhammad M. Rahman[11], Pulickel M. Ajayan[11*], Alan B. Dalton[4*], Manoj Tripathi[4*]

[1] CNR-Istituto Nanoscienze - Centro S3, Via Campi 213 41125 Modena, Italy

[2] École Polytechnique Fédérale de Lausanne (EPFL), Swiss Plasma Center (SPC), CH-5232 Villigen PSI, Switzerland.

[3] Department of Chemistry, Loughborough University, Loughborough LE11 3TU, United Kingdom

[4] Department of Physics and Astronomy, University of Sussex, Brighton BN1 9RH, U.K.

[5] Department of Physics, Informatics and Mathematics, University of Modena and Reggio Emilia, Via Campi 213 41125 Modena, Italy

[6] Otto Schott Institute of Materials Research, Friedrich Schiller University Jena, D-07743 Jena, Germany

[7] Fondazione Bruno Kessler Sensors and Devices, via Sommarive 18, 38123 Trento, Italy

[8] Laboratory of Bio-Inspired, Bionic, Nano, Meta, Materials & Mechanics, Department of Civil, Environmental and Mechanical Engineering, University of Trento, Via Mesiano, 77, 38123 Trento, Italy

[9] School of Engineering and Materials Science, Queen Mary University of London, Mile End Road, London E1 4NS, UK

[10] Department Civil and Environmental Engineering, South Dakota School of Mines and Technology, Rapid City, South Dakota 57701, USA

[11] Department of Materials Science and Nanoengineering, Rice University, Houston, TX 7705, USA

*Corresponding author

# Author contributed equally

**Abstract:** Friction-induced energy dissipation impedes the performance of nanoscale devices during their relative motion. Nevertheless, an ingeniously designed structure which utilises graphene topping can tune the friction force signal by inducing local strain. The present work reports capping of graphene over Si grooved surfaces of different pitch lengths from sub-nanoscale (P= 40 nm) to a quarter of a micron (P= 250 nm). The variation in the pitch lengths induces different strains in graphene revealed by scanning probe techniques, Raman spectroscopy and molecular dynamics (MD) simulation. The asymmetric straining of C-C bonds over the groove architecture is exploited through friction force microscopy in different directions of orthogonal and parallel to groove axis. The presence of graphene lubricates the textured surface by a factor of ≈10 and periodically dissipated friction force, which was found to be stochastic over the bare surface. For the first time, we presented transformation of the lubrication into an ultra-low friction force by a factor of ≈ 20 over the crest scanning parallel to the groove axis. Such anisotropy is found to be insignificant at the bare textured system, clearly demonstrating the strain-dependent regulation of friction force. Our results are applicable for graphene, and other 2D materials covered corrugated structures with movable components such as NEMS, nanoscale gears and robotics.

Engineered nanostructuring has significantly improved the sensing performance of Micro Electromechanical Systems (MEMS) and Nano Electromechanical Systems (NEMS)[1,2] devices by tuning wetting characteristics[3], nano-channeling[4], optical[5], mechanical[6] and electronic properties[7]. The requirement of nano/micro-machines has surged in modern-day technology, and the path is progressing towards miniaturized devices. On the other hand, nanoscale structured surfaces pose tremendous challenges of performance and efficiency when they contact one another (e.g. gear operation at nanoscale). The mass/force sensitivity and mechanical quality factors are inversely proportional to critical length scales[8]. The interaction forces, which are relatively weak at macro-scale (such as van der Waals and capillary) become dominant at the nano and sub-nano level. Therefore, nanostructured devices are often harmed under conditions of extreme temperature, pressure, friction, adhesion, and relative humidity[9]. The nano contacts exert enormous pressure at the interface even at low values of the applied normal force, subsequently leading to friction-induced wear. Thus, a novel strategy is needed to regulate the forces at the nanoscale, which could mitigate the physical alterations in the nanostructure under the effect of a high surface-to-volume ratio.

Several approaches have been adapted to tune the friction force by introducing liquid-state lubricants such as organic oils[10], ionic liquids[11], and tribological buffer layers such as polymer brushes[12]. Nevertheless, the ecologically harmful effect of liquid-state lubricants[13] and their inefficiency in confined conditions related to viscosity enhancements hinder their tribological performance at the nanoscale[10]. A valid alternative is represented by solid-state lubricants[14], in the form of nanoparticles or lamellar solids such as graphite and $MoS_2$. In the nanoscale regime, one of the most plausible solutions to protect surfaces relies on coating them with ultrathin lamellar sheets, i.e., the building blocks of lamellar solids, such as graphene monolayers. Graphene has the lowest bending rigidity[15,16], significantly high in-plane intrinsic strength[17] and is inert in humid and corrosive atmospheres[18]. However, the role of the substrate on which graphene is deposited is not passive and pivotal to modulate the mechanical, physical and electronic properties of the graphene film[19]. In particular, strain induced by interaction with the substrate is one of the intriguing parameters to adapt and tune graphene characteristics[20,21]. Recent works suggested the tribological characteristics of graphene could be modulated by strain or strain gradient fields[22–24]. Nevertheless, the frictional response of asymmetric strained graphene over the textured surface is a rarely addressed topic, which might play a crucial role in the durability of NEMS devices.

In the present study, we investigated the interplay between texture-induced strained graphene and its lubrication performance. We used nano-textured silicon surfaces as substrates that mimic the nano-gear and performed friction force microscopy (FFM) measurements in ambient condition. The different aspect ratios (pitch/depth) of the grooves modulate the conformation/suspension of graphene, resulting in its straining. Analytical modelling based on topography line profiles has been used to estimate strain values and adhesion energy between graphene and the textured surfaces. Raman spectroscopy has shown the systematic transformation of graphene compressive strain over flat surfaces which is released over textured regions as observed through softening of phonon modes. Molecular dynamics simulations corroborate the Raman measurements and exhibit atomic-scale resolution of graphene corrugation. The simulation results discern an asymmetric strain distribution through lattice expansion and contraction in C-C bond at different orientations. The strained graphene surface is further analyzed through FFM along the perpendicular and parallel directions relative to the groove axis, unravelling evident anisotropic frictional characteristics. The present work reveals for the first time regulation of the friction force dissipation in nanoscale architecture through strain engineering of graphene. The implementation of this study augments our understanding of superlattice structures, straintronics, and nanomachines.

**Results and discussion**

AFM images in **Figure 1(a, b, c)** show the morphology of bare Si (with native oxide) substrate at inset and graphene-covered textured surfaces referred as GrP40, GrP125 and GrP250. The preparation of the textured surface is described in previous work[25,26] and characterization in Supplementary Information **S1**. Each textured region comprises long parallel grooves separated by a variable distance (pitch length, P) of 40 ± 4, 125 ± 8 and 250 ± 14 nm. CVD produced polycrystalline graphene have been deposited by the wet transfer method. The topographic profiles of bare and covered surfaces, presented in **Figure 1 (d, e, and f),** illustrate graphene physical corrugation to the substrate. The measured depth of grooves is between 2.4-3 nm on bare P40, and is reduced by 10-15% after graphene deposition. On the other hand, the measured groove depths of bare P125 and P250 are ~ 4 nm and they are reduced by 7-10% in GrP125 and 3-5% GrP250, respectively, revealing different suspension depth of graphene over different textured surfaces. The conformation of graphene over the pattern is determined by the balance between interfacial adhesion and elastic energy stored in graphene (i.e., bending and stretching).[27,28]

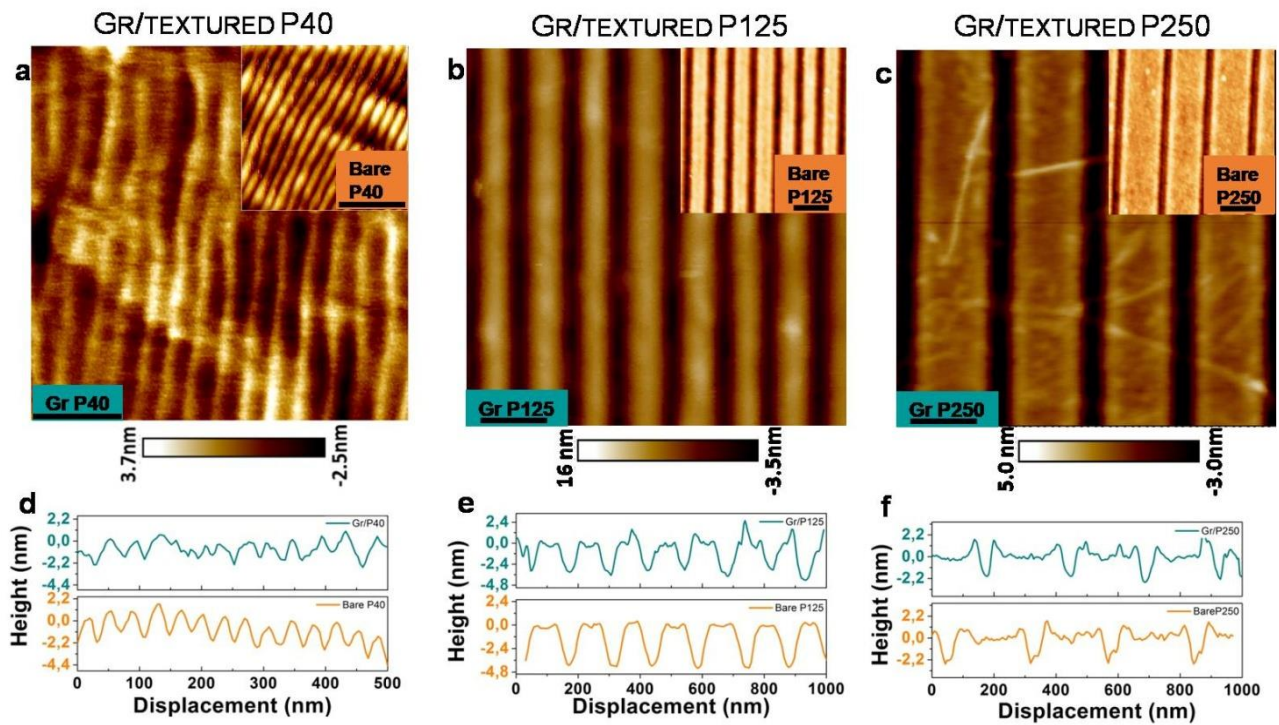

**Figure1: Morphology of graphene covered textured surfaces.** First row shows AFM topography of graphene covered textured surface of pitch (a) 40 ± 4 nm, (b) 125 ± 5 nm and (c) 250 ± 8 nm. Inset topographic images depict bare textured regions, Scale bars: 200 nm. The second row shows topographical line profiles of bare and graphene covered textured surfaces across the grooves for covered (dark cyan) and bare surface (orange) color.

The conformation induced average transversal strain ($\varepsilon$) (perpendicular to groove axis) and interfacial adhesion energy (meV Å$^{-2}$) between graphene and textured surfaces, estimated analytically from the height profiles, are reported in **Table 1** (see Supplementary Information **S2** for details). In this corrugation, suspension of graphene does not reach the stage of complete "snap-through"; nevertheless, a partial conformal contact is achieved[29]. Thus, graphene is considered as a membrane clamped between two grooves that induce different strains (see Supplementary **Figure S1**), which is increases with decreasing the P-value of the textured geometries. The analytical calculation also implies that tensile stress, deriving from the Poisson's ratio, should develop along the groove axis given by $\sigma_{parallel} = E\nu\varepsilon$, where $E$ is the Young's modulus and $\nu$ the Poisson's ratio. Therefore, the stretching of graphene perpendicularly to the groove axis also induces net tensile stress along its parallel axis, presented by MD simulation and exploited through FFM in the following sections.

**Table 1**: The interfacial interaction between graphene and textured surfaces of different pitch length.

|  | GrP(40) | GrP(125) | GrP(250) |
|---|---|---|---|
| **Conformation altitude (nm)** | 2.1 ± 0.30 | 3.5 ± 0.15 | 3.8 ± 0.05 |
| **Strain (%)** | ≈ 0.51 | ≈ 0.27 | ≈ 0.15 |
| **Interfacial adhesion energy (meV Å$^{-2}$)** | ≈ 0.48 | ≈ 0.16 | ≈ 0.11 |

The substrate induced stretching/compression of single layer graphene and the doping for each textured surface have been quantified by comparing Raman spectroscopy on the flat region (Gr/Flat) and graphene covered textured surfaces (Gr P40 to Gr P250). The Raman phonon modes of G peak position (PoSG) and 2D peak position (PoS2D) are associated with strain since a change in lattice constant leads to a variation in the phonon modes. Furthermore, these modes are useful to detect carrier concentration (n) due to alteration in bond length and non-adiabatic electronic-phonon coupling[31]. The relation between strain and doping of graphene with PoSG and PoS2D is described in Supplementary Information **S3.** It is known that wet chemical transferred graphene over flat Si substrate is a p-type doped system of compressive strain [32]. Thus, there is a release of the compressive strain in graphene at the textured region. We observed this phenomenon through gradual phonon softening in Raman modes of G and 2D of graphene deposited over Gr/Flat, GrP250, GrP125 and GrP40; see **Figure 2a, b**. Nevertheless, we did not observe the splitting of either G and 2D modes, which indicates that the magnitude of the induced strain is not high ($< \pm 0.35\%$)[33].

The correlation plot in **Figure 2c** shows the distribution of PoS2D as a function of PoSG, the mean value of the distribution being represented by star shapes. The strain axis (green color) and doping axis (brown color) are drawn at the slope ($\partial PoS2D/\partial PoSG$) range 2.25-2.8 and 0.75, respectively[34,35]. The intersection of the axis is assumed as a point of minimal strain and doping. The coordinates are taken from the work of Lee *et al.*[35] for laser line ($\lambda$ =514 nm); we introduced the modification for dispersive values of 2D ($\partial 2D/\partial E_{laser} \approx$ 88-110 cm$^{-1}$/eV)[36] for laser line ($\lambda$= 532 nm). The correlation plot illustrates relative change in the average compressive strain ($\varepsilon$) for Gr/flat is $\approx$-0.09%, which is transformed on textured surfaces as follows: P= 250 nm ($\varepsilon \approx$- 0.07%), P = 125 nm ($\varepsilon$ =- 0.061%), P = 40 nm ($\varepsilon$ = 0.02%). Unlike flat or biaxial strained surfaces (e.g. suspended graphene over a circular trench), the patterned surfaces induce anisotropy in graphene strained values oriented with groove axis[37]. Here, we report the integral of the strain values, which exhibit distinct variation along the perpendicular and parallel-to-groove axis. The correlation plot also deconvoluted the doping in graphene; a relative change in the carrier concentration (n) is observed from flat surface to GrP40 texture due to lesser coupling with the Si substrate. It has been found that the gap between graphene and Si substrate prohibits p-doping in graphene, as observed in the case of wrinkles showing lower work function[38].

It is worthy to note that a Raman laser spot diameter at 100X (objective lens) is nearly 700 nm x 700 nm. Therefore, the measured strain and the carrier concentration values are averaged over several crests, troughs and flat regions in each spectrum. Atomic-scale feature of graphene conformation over textured silicon surfaces have therefore been investigated through molecular dynamics (MD) and density functional theory (DFT) calculations of the graphene/Si at different pitch lengths, see supplementary **S4** for DFT and MD set-up. The crest region of GrP250 shows higher compressive strain induced through contact with the Si substrate, while the neighbouring trough exhibits curvature-induced tensile strain, which decreases with pitch length down to GrP40. The magnitude of net compression over a crest is proportional to the area of graphene in direct contact with the Si substrate, and is, therefore, higher than the tension across a trough, which leads to decreasing average values of compression, as shown in **Figure 2(d)**, which is excellent agreement with Raman spectroscopic measurements. A similar trend was observed by Zhang *et al.*[39] on biaxially strained graphene covered silicon nanospheres with different diameters. In that arrangement, the authors reported a change of compressive strain into tensile in graphene deposited over smaller spherical particles due to increasing real contact area at the apex. Hinnefeld *et al.*[34] found a similar trend for graphene suspended on silicon pillars with a separation distance of 600 nm showing an increase in charge carrier concentration and decreased compressive strain. Here, we bring down the textured spacing of one order of magnitude (i.e. ≈ 40 nm), where the deposited graphene resembles the characteristics of strain and doping of a partial suspended sheet, its net height variation is illustrated at inset **Figure 2(d)**. There is a generation of ripples in the suspended region as an act of releasing net compressive strain, which will be explored through FFM.

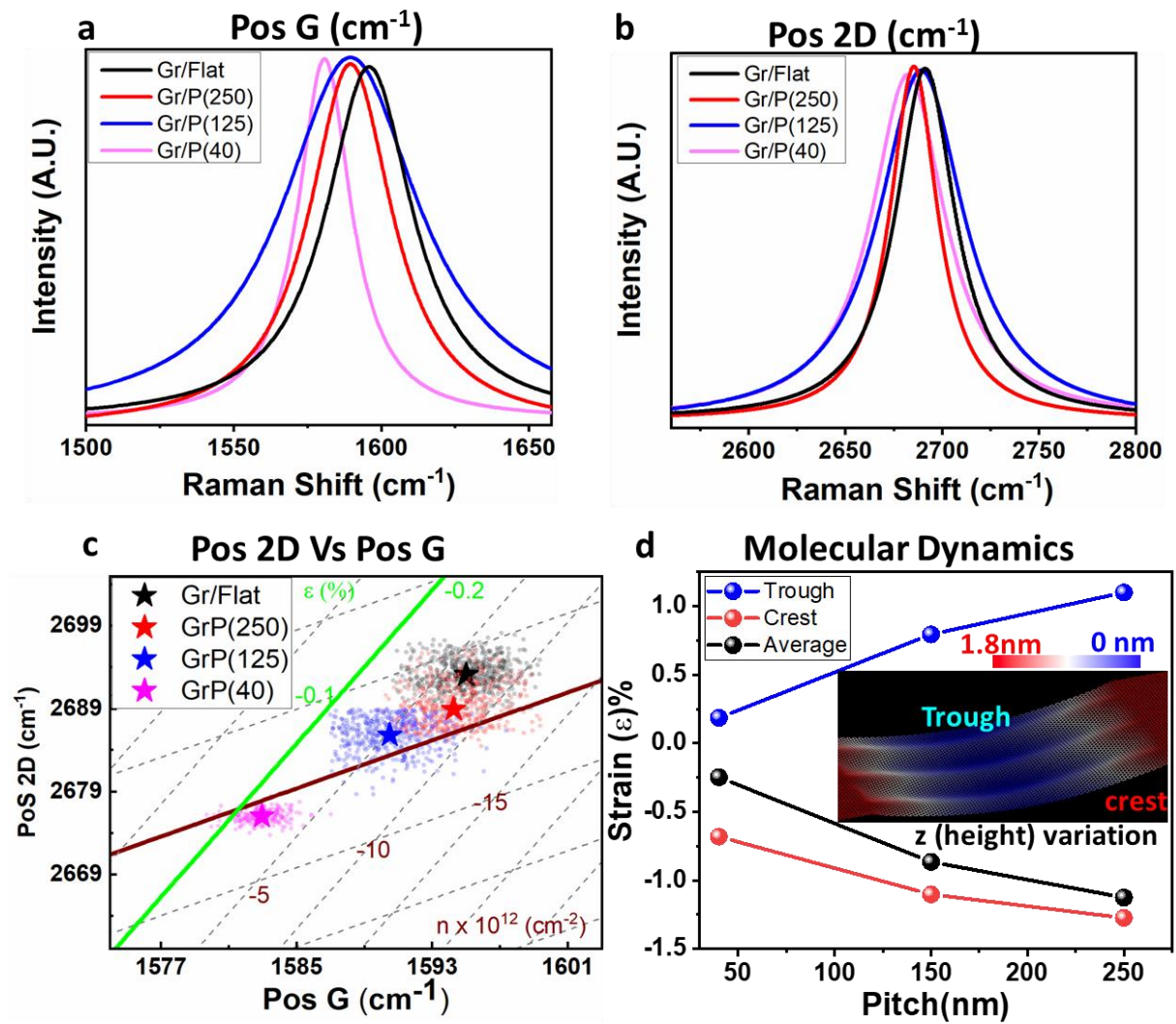

**Figure2: Raman spectrum of graphene covered textured region.** (a) Raman spectrum of $G_{pos}$(cm$^{-1}$) for graphene covered flat surface and different textured regions. (b) Raman spectrum of 2D(cm$^{-1}$) peak positions shows gradual red-shift from flat to the textured region. (c) Correlation plot of PosG vs Pos2D phonon modes for deconvoluted strain and doping in graphene from flat to the textured regions. The data distribution is from 70-100 spectra and the mean values are represented by star-shaped points. The strain (ε) axis is shown in green, and the brown color axis shows carrier concentration (n x$10^{12}$ cm$^{-2}$). (d) Bond strain distribution at the crest, trough and averaged over the entire surface (black color data) for different pitch lengths measured from MD simulations. Inset shows the net height (z-scale) variation at crest and trough regions for GrP40 under the influence of net tensile and compressive strain.

Raman spectroscopy revealed a significant disparity in strain values between Gr/Flat and GrP(40) configurations and thus, these extreme surfaces are chosen for FFM[40] measurements. Due to the intrinsic anisotropy in the textured induced strain in graphene, FFM measurements were performed in orthogonal **(Figure 3a-c)** and parallel **(Figure 3d-f)** directions of the groove axis of GrP(40) (details about procedure and calibration are reported in Supplementary Information **S5**). FFM images on GrP40 sample comprises bare textured silicon regions and nearby graphene covered areas in a single acquisition. This methodology allows comparing between bare and covered textured surfaces under similar contact conditions and local environment allows to disentangle the possible geometrical effects or tip shape contributions (see **Figure S4** for tip's curvature radius estimation). There is a significant contrast in the friction force values between bare and covered graphene for both orientations (**Figure 3b, e**), which shows the excellent lubrication performance of single-layer graphene over the highly periodic surface. The presence of graphene reduces average friction force up to 5-6 times than on bare surface under similar applied load conditions ranging from 10-30 nN, and no edge failure was noticed. These results are in agreement with previous nanotribological characterizations of graphene on flat silicon substrates[41–43]. The friction force profile in **Figure 3c** shows a markedly distinguishable undulated friction force response between graphene capped and bare silicon, orthogonal to the groove axis. The sliding tip results in a periodic modulation in friction force at the graphene capped region, which is a relatively irregular distribution over bare silicon. The periodic friction force modulation might be achieved by pinning the sliding tip and then a sudden jump to the next minima driven by the cantilever movement and favoured by the slippery crest.

While scanning along parallel to the groove axis at the capped region, friction force modulation as a function of tip displacement is practically absent, though stochastic friction force is sustained at the bare surface. It is clearly illustrated in the friction force map in **Figure 3e,** and it is associated with the profile drawn orthogonal to the grooves axis (**Figure 3f**) to provide a valid comparison with **Figure 3c**. The detailed analysis between crest and trough for the scanned orthogonal and parallel revealed a remarkable difference (Gr/LF$_{\text{Trough parallel}}$ -Gr/LF$_{\text{crest parallel}}$ ) ≈ 0.2nN and (Gr/LF$_{\text{Trough orthogonal}}$ -Gr/LF$_{\text{crest orthogonal}}$ ) ≈ 1.5nN, which is a nearly sevenfold increase. The ratio of the friction force at trough/crest measured during the scan in parallel and orthogonal directions at fixed load conditions is ≈ 2 and 5, respectively. Thus, the trough region of an orthogonally scanned textured surface contributing to the highest friction force, which is suppressed along the parallel scanned region. Nevertheless, we observed the isotropic frictional response over the bare Si textured surface scanned in orthogonal and parallel directions, as expected for this design of texturing[45]. It indicates that the strained components in graphene play a pivotal role in regulating the friction force induced from the textured surface.

While scanning parallel over the crest, the tip motion direction is perpendicular to the strain axis induced from a groove. This favours smaller friction force and results in super-lubricity, distinct to slide parallel to the strain axis. This geometrically induced anisotropic response in the friction force allows its tuning at the nanoscale. Therefore, a desired friction force could be achieved with the textured orientation useful to regulate several sliding of nanoscale objects, designer diffusion gradient for add molecules or even manifestation of biological cells proliferation. Such a high degree of friction force regulation is not possible over the graphene covered flat surfaces, which shows similar friction force (isotropic) in different scanning directions, see Supplementary Information (**Figure S5**). Though, anisotropy in friction due to different arrangement of carbon atoms in graphene cannot be neglected[46,47]. Here, we display the friction force regulation through graphene covered textured surface as an alternative platform, which is relatively easy to achieve than the controlling atomic arrangement of carbon atoms.

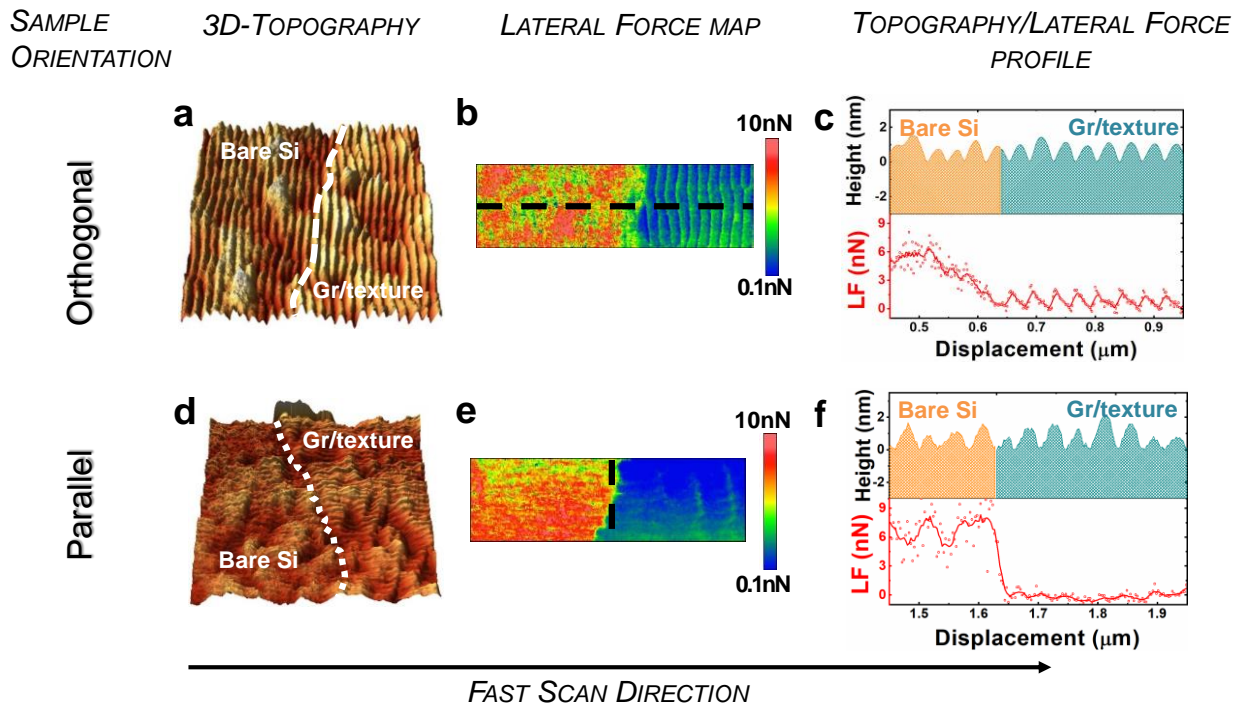

**Figure 3. Effect of scan direction on friction force for GrP(40).** First row shows (a) topography image (1.0 x 1.0 micron) and (b) friction force map (1.0 x 0.3 micron) measured at applied normal load ≈30nN simultaneously on P40 for grooves axis aligned orthogonal to the fast scan direction. The white dashed line in topography profiles represents the interface between the bare and graphene covered region. (c) Top, height profile (orange color correspond to bare silicon, dark cyan color to graphene covered region) and, bottom, corresponding friction force profile (open dots correspond to raw data distribution, the red line is the result of mean values extracted from black dashed line in (b). Second row shows (d) topography image (1.0 x 1.0 micron) and (e) friction force map (1.0 x 0.3 micron) or groove axis aligned parallel to the fast scan direction at applied normal load ≈25nN. (f) Top, height profile and bottom, corresponding friction force profile extracted from black dashed line in (e).

The load dependence friction curves for P(40) and GrP(40) are reported in **Figure 4** for orthogonal and parallel scanned directions (see details in Supplementary Information **S6**, **Figure S6-S8**). The friction force values for the bare textured region are increased by a factor of eight as compared to graphene- covered regions for all applied loads range (-10 to 30 nN), consistent with friction force profile. The shear strength (S = friction force/area) of the interface is measured by fitting the data through Derjaguin-Muller-Toporov (DMT) model (continuous line in **Figure 4**) following 2/3 power law within continuum mechanical modelling of the contact region[48–50] and COF is measured by a linear fit of the curves (dashed lines). The results are shown in **Table 2,** revealing a nearly three times change in the S for GrP(40) region between parallel and orthogonal to the groove axis. While comparable (only 1.07 times) S is measured for bare P(40), scanned between parallel and orthogonal directions. Also,the S measured for sliding parallel to the groove on GrP40 are lower by ≈ 50% than Gr/Flat, which is in good agreement with the literature [42,51]. The COF values are corroborated with S revealing minimal values of 0.009±0.001 and 0.011±0.002 for different locations. The COF values for the Gr/Flat surface was found to be intermediate between orthogonal and parallel scanned axis as follows: Gr/parallel < Gr/Flat < Gr/perpendicular. The present work is in good agreement with the investigation demonstrated by Zhang and coworkers[52] for tuning COF by regulating strain in the suspended graphene. The COF of the suspended graphene (a region of low strain) is almost double to the strained (0.3%) graphene. The presented textured surfaces demonstrate that crest and trough serve as distinct strained regions that can regulate the friction force. The FFM values for Gr/Flat represent a compressive strain system, as demonstrated in the Raman correlation plot. Here, elastic buckle formation as a "puckering effect" in front of sliding tip apex leads to the higher friction force values[53].

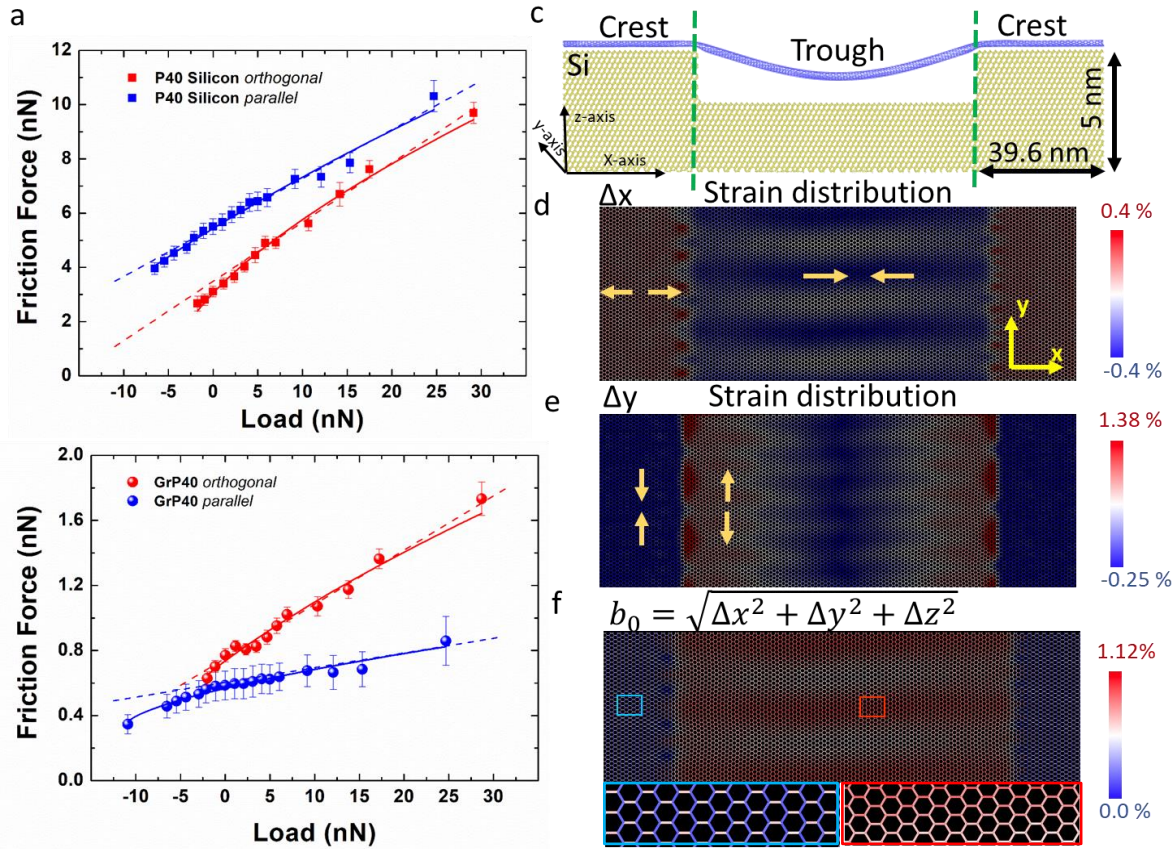

**Figure 4: Friction Force as a function of load applied in parallel and perpendicular to groove axis:** Load dependent friction force curves on GrP40 sample with the grooves axis oriented parallel (blue) and perpendicular (red) to the fast scan direction; (a) on bare silicon textured surface (b) on adjacent graphene covered region. Triangular and circular shaped data represent the experimental values, continuous lines are the fitting curve from DMT model and dashed line is the linear fit. (c) MD simulation of a graphene sheet sags into the P40 textured Si surface. The vertical drawn dashed green lines represent the trough region of suspended graphene between two crests. (d) Strain distribution based on bond strain variation along the x-axis ($\Delta x$), (e) y-axis ($\Delta y$) and (f) total bond length ($b_0$). The Inset region (marked by the colored rectangle in panel (f)) shows the variation in C-C bond length in the crest and trough regions. The asymmetry in $b_0$ between different regions and along different axes is readily apparent, as shown in the zoom-in image.

The anisotropic values of friction force, consequently S and COF for the graphene covered textured surface can be explained through different stretching/compression of C-C bonds in either direction over an individual groove. It has been validated through MD simulation for graphene over P40 architecture at **Figure 4 (c)**. The strain distribution in graphene along the crest and trough (between green dashed lines in panel (c)) in orthogonal (x), parallel (y) and out-of-plane to the silicon surface has been calculated through percentage changes in Δx, Δy and bond length $b_o = \sqrt{\Delta x^2 + \Delta y^2 + \Delta z^2}$ with respect to Gr/Flat (for more details, see Supplementary Information S4) respectively, and is shown in **Figure (d, e, f)**. Along the x-axis, the bending of graphene at the crest-trough interface generates localized tensile strained regions (red color). This stretching of C-C bonds at the interface leads to net compressive strain at the trough. Along the y-axis, there is a net tensile strain at the trough region and compressive over the crest resulting from the combined effect of substrate adhesion and adjacent suspended graphene, see **Figure 4(e)**; although the magnitude of this compression is smaller than in Δx strain distribution. The integral bond length ($b_o$) distribution at the crest illustrates asymmetric bond alteration along orthogonal (stretching) and parallel (compressive) to groove axis. This asymmetry is also sustained at the trough, but a higher magnitude has been observed (see **Figure 4(f))** and its inset marked by rectangles). This asymmetric bond length distribution results in anisotropy in friction forces orthogonal and parallel to the groove axis. Also, higher C-C bond elongation at the trough region in the x-axis (red strips) results in a net compressive strain in the y-axis released through the formation of ripples, see inset **Figure 2(d)**. This explains the higher friction force over the trough region, as the presence of ripple favours the puckering effect[47], while sliding direction parallel to the groove axis. It clearly shows the remarkable tribological (friction force, COF, S) performance of graphene for wide anisotropy over the same textured surface, which was not possible over the traditional Gr/Flat. It is indeed the motivation for further research to explore tailored friction values from directional strained graphene. Thus, graphene covered textured systems could bring an era of tuned friction force at the nanoscale, which has been a non-trivial task in the last decades. Moreover, such regulated friction could enhance the performance of nanomachines.

**Table 2**: Shear strength (MPa) and coefficient of friction (COF) from load-dependent friction force curve.

| SAMPLE | SAMPLE ORIENTATION | SHEAR STRENGTH | | FRICTION COEFFICIENT | |
|---|---|---|---|---|---|
| | | Location 1 | Location 2 | Location 1 | Location 2 |
| GrP40 | Orthogonal | 38 MPa | 47 MPa | 0.033±0.002 | 0.040±0.002 |
| GrP40 | Parallel | 12 MPa | 17 MPa | 0.009±0.001 | 0.011±0.002 |
| P40 Silicon | Orthogonal | 345 MPa | | 0.21±0.01 | |
| P40 Silicon | Parallel | 322 MPa | | 0.18±0.01 | |
| Gr- Flat | | 25 MPa | | 0.025±0.003 | |

## Conclusion

In summary, the graphene over textured surfaces can offer a wide range of opportunities due to the interplay between adhesion force energy, bending, stretching, and strained orientation. By controlling the grooves separation distance, a tunable strain in a single layer of graphene can be achieved, presented through the analytical model, MD simulation and Raman spectroscopic measurements. The graphene deposited over flat Si surface revealed a compressively strained system that systematically releases over the textured surfaces. The overlayered graphene drops the friction force values and channelizes the friction dissipation, complementary to the textured geometry. The strain distribution in graphene over the textured architecture regulates the friction force; consequently, COF and S values. Therefore, single-layer graphene deposited onto an anisotropic nanotextured system could acquire diverse nanomechanical properties. It is demonstrated in reference to the FFM that depends on the sliding direction with respect to grooves orientation. The presented work will pave the pathway to nanoscale devices for efficient functioning and controlled motion of nanoscale objects, emphasizing nanomechanics and nanorobotics.

## Materials and Methods Section

Preparation of textured surface: Texture of the P(40) sample was obtained by low energy ion irradiation of a (100) silicon wafer over $1 \times 1$ cm$^2$ area. Ion beam parameters (1 keV O$^+$ ion beams, with fluxes $1.1 \times 10^{13}$ cm$^{-2}$ s$^{-1}$ and angle of incidence of 50°) were optimized to obtain a texture of quasi parallel grooves with nanometer height. Texture of the P(125) and P(250) samples was obtained by focused ion beam (FIB) milling of a (001) silicon wafer (P doped, resistivity 500−3000 Ω cm). Milling was performed with a dual beam FIB apparatus (FEIDB235M) using a 30KeV Ga$^+$ ion beam at normal incidence with ion dose of $2.25 \times 10^{16}$ ions/cm$^2$. The resulting structure were square-textured areas (10μm $\times$ 10μm) composed of parallel nano-grooves about 50 nm wide with variable pitches of 125 and 250 nm. The optimization of FIB milling procedure have been previously published by some of the authors and readers are referred to this publication for further details[25]. The AFM characterization is reported in supplementar Information **S1**.

Graphene deposition and cleaning: Commercial single layer CVD graphene by ACS Material (Pasadena, CA-USA) and Graphenea Inc. (Spain) was deposited on nanostructure surfaces through the standard method of polymer assisted wet transfer followed by removal of polymer residue in an acetone bath (400°C for 30 min). Later, samples were dried in oven at 400 C for 20 min and eventually heated in vacuum at 300°C for 2 hrs. Mechanical cleaning have been implemented by sacrificial AFM tip prior to the friction measurements.

Raman measurements: Raman Spectroscopy (spectral resolution 0.3 cm-1 from Renishaw) has been performed using 100X objective lens. The Power of the LASER used was 10% (through ND filters) for the exposure of 5seconds using 1800 grating. The Raman modes of G and 2D peak are fitted with Lorentzian curve to evaluate the peak positions and its spectral shift (cm$^{-1}$).

Atomic force microscopy and Friction Force Microscopy: Two different Atomic Force Microscope (AFM) were utilized during the experiments. The Bruker Dimension Icon with Peak Force Tapping Mode$^{TM}$ option and the NT-MDT NTEGRA AURA system. All the measurements were carried out in air, under ambient conditions. Commercially available rectangular shaped silicon cantilevers (MikroMaschHQ : CSC37/NoAl) with nominal normal elastic constants between 0.2 and 0.8 N m$^{-1}$ were used for Friction Force Microscopy (FFM) measurements. The detailed calibration procedure for the measurements is mentioned in supplementary **S5**.

**Acknowledgement:**

MT and ABD would like to acknowledge strategic development funding from the University of Sussex. JGM acknowledges the use of the HPCMidlands + facility, funded by EPSRC grant EP/P020232/1 as part of the HPC Midlands+ consortium. PMA acknowledges support from the Air Force Office of Scientific Research under award number FA9550-18-1-0072. A.M. and G.P. like to acknowledge support from MIUR, PRIN 2017 project n.2017PZCB5 – UTFROM; A.R. G.P. and S.V. like to acknowledge support from Regione Emilia Romagna, Project INTERMECH and Project n. PG/2018//631311-RIMMEL. NMP has received funding from the European Union's Horizon 2020 Research and Innovation Programme under grant agreement GrapheneCore3 n. 881603. We thank G.Gazzadi (CNR-Istituto Nanoscienze) for P125 and P250 substrate sculpting by Focused Ion Beam

**References**

[1] A. Balčytis, J. Juodkazytė, G. Seniutinas, X. Li, G. Niaura, S. Juodkazis, in *Laser-Based Micro- Nanoprocessing X* (Eds.: U. Klotzbach, K. Washio, C.B. Arnold), **2016**, p. 97360G.

[2] X. Zang, Q. Zhou, J. Chang, Y. Liu, L. Lin, *Microelectron. Eng.* **2015**, *132*, 192.

[3] J. Bico, U. Thiele, D. Quéré, *Colloids Surfaces A Physicochem. Eng. Asp.* **2002**, *206*, 41.

[4] C. Cottin-Bizonne, J.-L. Barrat, L. Bocquet, E. Charlaix, *Nat. Mater.* **2003**, *2*, 237.

[5] L. A. Dobrzański, A. Drygała, K. Gołombek, P. Panek, E. Bielańska, P. Zieba, *J. Mater. Process. Technol.* **2008**, *201*, 291.

[6] K. J. Hemker, W. N. Sharpe Jr, *Annu. Rev. Mater. Res.* **2007**, *37*, 93.

[7] S. Finkbeiner, in *2013 Proc. ESSCIRC*, IEEE, **2013**, pp. 9–14.

[8] S. Mukherjee, N. R. Aluru, **2006**.

[9]  Y. Ping, L. N. Bo, Y. Daoguo, L. J. Ernst, *Microsyst. Technol.* **2006**, *12*, 1125.

[10] J. Klein, E. Kumacheva, *Science (80-. ).* **1995**, *269*, 816.

[11] O. Y. Fajardo, F. Bresme, A. A. Kornyshev, M. Urbakh, *Sci. Rep.* **2015**, *5*, 7698.

[12] S. Lee, N. D. Spencer, *Science (80-. ).* **2008**, *319*, 575.

[13] O. Hod, E. Meyer, Q. Zheng, M. Urbakh, *Nature* **2018**, *563*, 485.

[14] A. Rosenkranz, H. L. Costa, M. Z. Baykara, A. Martini, *Tribol. Int.* **2021**, *155*, 106792.

[15] D.-B. Zhang, E. Akatyeva, T. Dumitrică, *Phys. Rev. Lett.* **2011**, *106*, 255503.

[16] H. Chen, S. Chen, *J. Phys. D. Appl. Phys.* **2013**, *46*, 435305.

[17] M. R. Vazirisereshk, H. Ye, Z. Ye, A. Otero-de-la-Roza, M.-Q. Zhao, Z. Gao, A. T. C. Johnson, E. R. Johnson, R. W. Carpick, A. Martini, *Nano Lett.* **2019**, *19*, 5496.

[18] G. Chilkoor, N. Shrestha, A. Kutana, M. Tripathi, F. C. Robles Hernández, B. I. Yakobson, M. Meyyappan, A. B. Dalton, P. M. Ajayan, M. M. Rahman, V. Gadhamshetty, *ACS Nano* **2021**, 3987.

[19] M. Tripathi, F. Lee, A. Michail, D. Anestopoulos, J. G. McHugh, S. P. Ogilvie, M. J. Large, A. A. Graf, P. J. Lynch, J. Parthenios, K. Papagelis, S. Roy, M. A. S. R. Saadi, M. M. Rahman, N. M. Pugno, A. A. K. King, P. M. Ajayan, A. B. Dalton, *ACS Nano* **2021**, *15*, 2520.

[20] A. M. van der Zande, R. A. Barton, J. S. Alden, C. S. Ruiz-Vargas, W. S. Whitney, P. H. Q. Pham, J. Park, J. M. Parpia, H. G. Craighead, P. L. McEuen, *Nano Lett.* **2010**, *10*, 4869.

[21] N. Levy, S. A. Burke, K. L. Meaker, M. Panlasigui, A. Zettl, F. Guinea, A. H. C. Neto, M. F. Crommie, *Science (80-. ).* **2010**, *329*, 544.

[22] X. Wang, K. Tantiwanichapan, J. W. Christopher, R. Paiella, A. K. Swan, *Nano Lett.* **2015**, *15*, 5969.

[23] C. Wang, S. Chen, *Sci. Rep.* **2015**, *5*, 1.

[24] C. Dai, Z. Guo, H. Zhang, T. Chang, *Nanoscale* **2016**, *8*, 14406.

[25] A. Rota, M. Tripathi, G. Gazzadi, S. Valeri, *Langmuir* **2013**, *29*, 5286.

[26] R. Dell'Anna, E. Iacob, M. Tripathi, A. Dalton, R. Böttger, G. Pepponi, *J. Microsc.* **2020**, jmi. 12908.

[27] J. S. Bunch, M. L. Dunn, *Solid State Commun.* **2012**, *152*, 1359.

[28] W. G. Cullen, M. Yamamoto, K. M. Burson, J.-H. Chen, C. Jang, L. Li, M. S. Fuhrer, E. D.

Williams, *Phys. Rev. Lett.* **2010**, *105*, 215504.

[29] T. J. W. Wagner, D. Vella, *Appl. Phys. Lett.* **2012**, *100*, DOI 10.1063/1.4724329.

[30] M. B. Elinski, Z. Liu, J. C. Spear, J. D. Batteas, *J. Phys. D. Appl. Phys.* **2017**, *50*, DOI 10.1088/1361-6463/aa58d6.

[31] M. Lazzeri, F. Mauri, *Phys. Rev. Lett.* **2006**, *97*, 266407.

[32] S. Goniszewski, M. Adabi, O. Shaforost, S. M. Hanham, L. Hao, N. Klein, *Sci. Rep.* **2016**, *6*, 22858.

[33] C. Kong, C. Pilger, H. Hachmeister, X. Wei, T. H. Cheung, C. S. W. Lai, N. P. Lee, K. K. Tsia, K. K. Y. Wong, T. Huser, *Light Sci. Appl.* **2020**, *9*, DOI 10.1038/s41377-020-0259-2.

[34] J. H. Hinnefeld, S. T. Gill, N. Mason, *Appl. Phys. Lett.* **2018**, *112*, 173504.

[35] J. E. Lee, G. Ahn, J. Shim, Y. S. Lee, S. Ryu, *Nat. Commun.* **2012**, *3*, 1024.

[36] L. M. Malard, M. A. A. Pimenta, G. Dresselhaus, M. S. Dresselhaus, *Phys. Rep.* **2009**, *473*, 51.

[37] J. K. Lee, S. Yamazaki, H. Yun, J. Park, G. P. Kennedy, G. T. Kim, O. Pietzsch, R. Wiesendanger, S. Lee, S. Hong, U. Dettlaff-Weglikowska, S. Roth, *Nano Lett.* **2013**, *13*, 3494.

[38] F. Long, P. Yasaei, R. Sanoj, W. Yao, P. Král, A. Salehi-Khojin, R. Shahbazian-Yassar, *ACS Appl. Mater. Interfaces* **2016**, *8*, 18360.

[39] Y. Zhang, M. Heiranian, B. Janicek, Z. Budrikis, S. Zapperi, P. Y. Huang, H. T. Johnson, N. R. Aluru, J. W. Lyding, N. Mason, *Nano Lett.* **2018**, *18*, 2098.

[40] E. Gnecco, R. Pawlak, M. Kisiel, T. Glatzel, E. Meyer, in *Nanotribology and Nanomechanics*, Springer International Publishing, Cham, **2017**, pp. 519–548.

[41] S. Zhang, T. Ma, A. Erdemir, Q. Li, *Mater. Today* **2019**, *26*, 67.

[42] Z. Deng, N. N. Klimov, S. D. Solares, T. Li, H. Xu, R. J. Cannara, *Langmuir* **2013**, *29*, 235.

[43] N. Manini, G. Mistura, G. Paolicelli, E. Tosatti, A. Vanossi, *Adv. Phys. X* **2017**, *2*, 569.

[44] H. Li, A. W. Contryman, X. Qian, S. M. Ardakani, Y. Gong, X. Wang, J. M. Weisse, C. H. Lee, J. Zhao, P. M. Ajayan, *Nat. Commun.* **2015**, *6*, 1.

[45] A. Berardo, G. Costagliola, S. Ghio, M. Boscardin, F. Bosia, N. M. Pugno, *Mater. Des.* **2019**, *181*, 107930.

[46] C. M. Almeida, R. Prioli, B. Fragneaud, L. G. Cançado, R. Paupitz, D. S. Galvão, M. De

Cicco, M. G. Menezes, C. A. Achete, R. B. Capaz, *Sci. Rep.* **2016**, *6*, 1.

[47] J. S. Choi, J.-S. Kim, I.-S. Byun, D. H. Lee, M. J. Lee, B. H. Park, C. Lee, D. Yoon, H. Cheong, K. H. Lee, Y.-W. Son, J. Y. Park, M. Salmeron, *Science (80-. ).* **2011**, *333*, 607.

[48] R. W. Carpick, D. F. Ogletree, M. Salmeron, *Area* **1999**, *400*, 395.

[49] U. D. Schwarz, O. Zwörner, P. Köster, R. Wiesendanger, *Phys. Rev. B - Condens. Matter Mater. Phys.* **1997**, *56*, 6997.

[50] T. D. B. Jacobs, A. Martini, *Appl. Mech. Rev.* **2017**, *69*, DOI 10.1115/1.4038130.

[51] R. Buzio, A. Gerbi, S. Uttiya, C. Bernini, A. E. Del Rio Castillo, F. Palazon, A. S. Siri, V. Pellegrini, L. Pellegrino, F. Bonaccorso, *Nanoscale* **2017**, *9*, 7612.

[52] S. Zhang, Y. Hou, S. Li, L. Liu, Z. Zhang, X.-Q. Feng, Q. Li, *Proc. Natl. Acad. Sci.* **2019**, *116*, 24452.

[53] C. Lee, Q. Li, W. Kalb, X. Z. Liu, H. Berger, R. W. Carpick, J. Hone, *Science (80-. ).* **2010**, *328*, 76.

# Supporting Information

# Anisotropic frictional response of strained graphene

Andrea Mescola[1#], Guido Paolicelli[1#], Roberto Guarino[2], James G. McHugh[3], Sean P. Ogilvie[4], Alberto Rota[1,5], Enrico Gnecco[6], Erica Iacob[7], Sergio Valeri[1,5], Nicola M. Pugno[8,9], Venkata Gadhamshetty[10], Muhammad M. Rahman[11], Pulickel M. Ajayan[11*], Alan B. Dalton[4*], Manoj Tripathi[4*]

[1] CNR-Istituto Nanoscienze - Centro S3, Via Campi 213 41125 Modena, Italy

[2] École Polytechnique Fédérale de Lausanne (EPFL), Swiss Plasma Center (SPC), CH-5232 Villigen PSI, Switzerland.

[3] Department of Chemistry, Loughborough University, Loughborough LE11 3TU, United Kingdom

[4] Department of Physics and Astronomy, University of Sussex, Brighton BN1 9RH, U.K.

[5] Department of Physics, Informatics and Mathematics, University of Modena and Reggio Emilia, Via Campi 213 41125 Modena, Italy

[6] Otto Schott Institute of Materials Research, Friedrich Schiller University Jena, D-07743 Jena, Germany

[7] Fondazione Bruno Kessler Sensors and Devices, via Sommarive 18, 38123 Trento, Italy

[8] Laboratory of Bio-Inspired, Bionic, Nano, Meta, Materials & Mechanics, Department of Civil, Environmental and Mechanical Engineering, University of Trento, Via Mesiano, 77, 38123 Trento, Italy

[9] School of Engineering and Materials Science, Queen Mary University of London, Mile End Road, London E1 4NS, UK

[10] Department Civil and Environmental Engineering, South Dakota School of Mines and Technology, Rapid City, South Dakota 57701, USA

[11] Department of Materials Science and Nanoengineering, Rice University, Houston, TX 7705, USA

*Corresponding author

# Author contributed equally

1. **Sample Preparation**
2. **Graphene deposition and cleaning**
3. **Estimation of strain and adhesion energy**
4. **Raman Measurement**
5. **Molecular Dynamics calculation**
6. **Calculation of strain and doping from Raman modes in Graphene**
7. **AFM and Tribology**
8. **Friction versus load curves fitting procedure**

## S1. Morphology Characterization and conformation

P(40) sample

Sample was characterized by AFM using semi-contact mode imaging in air. The period of the texture structure along the direction of incident ions was calculated as the first maximum of height-height Auto Correlation Function (ACF), and it result about $40 \pm 2$ nm. The height of the grooves $h$ i.e. the average difference between peak heights and valley depths of the grooves, was calculated as the maximum vertical distances between the highest and lowest data points within one period and for all the possible periods contained in a single row. The result of this analysis indicates $h_b = 2.8 \pm 0.6$ nm for the bare sample and $h_{gr} = 2.4 \pm 0.7$ nm for the graphene covered sample i.e. a relative difference of 15 %. The optimization of low energy ion irradiation procedure as well as the characterization methods have been comprehensively discussed elsewhere [1,2].

## S2. Estimation of strain and adhesion energy

Let us assume the profile of the substrate and the deformed shape of the graphene sheet can be approximated by continuous periodic functions, respectively $y_s(x)$ and $y_G(x)$.

Considering only one period, the graphene sheet can be assumed to have a length equal to the wavelength $\lambda$ in the undeformed shape (i.e., it is completely flat before the interaction with the substrate). After deformation, the new length of the graphene sheet in the x-y plane is given by the line integral:

$$l' = \int_L y_G(x)\, dx = \int_0^\lambda \sqrt{1 + \left(\frac{\partial y_G}{\partial x}\right)^2}\, dx \tag{1}$$

Consequently, the average strain of the graphene sheet due to in-plane stretching can be computed as:

$$\varepsilon_x = \frac{l' - \lambda}{\lambda} \tag{2}$$

With regards to the adhesion energy, in order to simplify the calculation and obtain a realistic order-of-magnitude estimation, let us consider the schematic profiles of the substrate and the graphene sheet shown in the Figure below. The parameters $\delta_1$, $\delta_2$ and $\alpha$ define the regions of the graphene sheets that adhere to the substrate (i.e., that are not suspended). With these assumptions, Equation (2) becomes:

$$\varepsilon_x \approx \frac{2\delta_2}{\lambda}(1 - \cos\alpha) \tag{3}$$

and, consequently, the total stretching energy is given by:

$$\Phi_m = \frac{1}{2} E t \lambda b \varepsilon_x^2 \approx \frac{2 E t b}{\lambda} \delta_2^2 (1 - \cos\alpha)^2 \tag{4}$$

being $E$ the in-plane Young's modulus of the graphene sheet, $t$ its thickness and $b$ its width.

The total adhesion energy, instead, is given by:

$$U_{adh} = -4\gamma b(\delta_1 + \delta_2) \tag{5}$$

where $\gamma$ is the adhesion energy per unit area.

Finally, from the condition:

$$\tag{6}$$

$$\frac{\partial(\Phi_m + U_{adh})}{\partial \delta_2} = 0$$

we obtain the adhesion energy per unit area, i.e.:

$$\gamma \approx \frac{Et}{\lambda}(1 - \cos\alpha)^2 \delta_2 \qquad (7)$$

The Table below reports the results of the estimation of the strain and of the adhesion energy for the three considered profiles, where we have used $E = 1$ TPa and $t = 0.34$ nm.

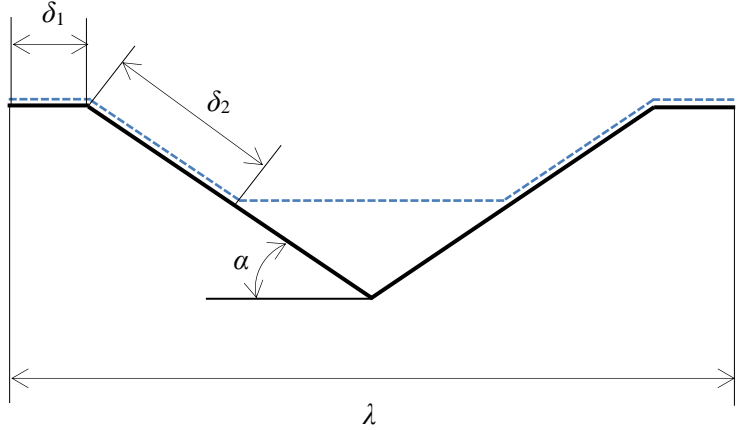

**Figure S1**: Schematic illustration of sagging graphene in the trough region between two grooves. λ (nm) is the spacing between the grooves of flat graphene. The parameters $\delta_1$, $\delta_2$ and $\alpha$ are associated to the adhesion of graphene between the grooves.

Table 1 below reports the results of the estimation of the average strain and adhesion energy for P40, P125 and P250

**Table S 1: Quantitative evaluation/ Estimation of conformation induced average strain and adhesion energy for GrP40, P125 and P250.**

|  | **GrP40** | **GrP125** | **GrP250** |
|---|---|---|---|
| **function** | $y_G(x) = A\sin\left(\frac{2\pi}{\lambda}x\right)$ | $y_G(x) = A_1 \sin^2\left(\frac{\pi}{\lambda}x\right) + A_2 \sin^2\left(\frac{2\pi}{\lambda}x\right)$ | $y_G(x) = A_1\sin(B_1 x + C_1) + A_2\sin(B_2 x + C_2) + A_3\sin(B_3 x + C_3)$ |
| **wavelength** | λ = 40 nm | λ = 125 nm | λ = 270 nm |
| **fit parameters** | $A \approx 0.77$ nm | $A_1 \approx 2.85$ nm<br>$A_2 \approx 1.56$ nm | $A_1 \approx 0.6962$ nm<br>$B_1 \approx 0.06903$ nm$^{-1}$<br>$C_1 \approx -0.9578$<br>$A_2 \approx 0.545$ nm<br>$B_2 \approx 0.04717$ nm$^{-1}$<br>$C_2 \approx 2.663$<br>$A_3 \approx 0.5382$ nm |

| | | | $B_3 \approx 0.09367$ nm$^{-1}$<br>$C_3 \approx 1.141$ |
|---|---|---|---|
| **average strain** | $\varepsilon \approx 0.33$ % | $\varepsilon \approx 0.28$ % | $\varepsilon \approx 0.14$ % |
| **adhesion parameters** | $\delta_1 \approx 0$ nm<br>$\delta_2 \approx 11.7$ nm<br>$\alpha \approx 7.6°$ | $\delta_1 \approx 25$ nm<br>$\delta_2 \approx 30.2$ nm<br>$\alpha \approx 6.1°$ | $\delta_1 \approx 105$ nm<br>$\delta_2 \approx 30.2$ nm<br>$\alpha \approx 6.7°$ |
| **adhesion energy** | $\gamma \approx 0.0073$ J m$^{-2} \approx$<br>$\approx 0.45$ meV Å$^{-2}$ | $\gamma \approx 0.0026$ J m$^{-2} \approx$<br>$\approx 0.16$ meV Å$^{-2}$ | $\gamma \approx 0.0017$ J m$^{-2} \approx$<br>$\approx 0.11$ meV Å$^{-2}$ |

We observe, as expected, that the average strain decreases for an increasing pitch of the texture substrate. This is due to the fact that the graphene sheet is not constrained between closer peaks, as instead happens for the 40 nm pitch. For the 250 nm pitch, in fact, the graphene sheet adheres completely to the substrate. We observed that the estimated adhesion energy per unit area is always in the same order of magnitude but it shows a decreasing trend with increasing wavelength of the profile. Note that the ion implantation process has heavily modified the SiO$_2$ surface, thus the order-of-magnitude of estimated adhesion energy is reasonable with values available in literature. For instance, Sabio *et al.*[3] estimated an interaction energy of 0.4 meV Å$^{-2}$ for graphene on SiO$_2$ on the basis of electrostatic interactions. Aitken and Huang showed that the effective adhesion energy of monolayer graphene on an oxide substrate also depend on the surface corrugation[4].

**S3. Calculation of strain and doping from Raman modes in Graphene**

The strain *(ε)* and the charge carrier concentration *(n)* of graphene are related to the Raman shift (ω$_1$, ω$_2$) as presented in equation (1)[5,6].

$$\begin{pmatrix} \omega_1 \\ \omega_2 \end{pmatrix} = T \begin{pmatrix} \varepsilon \\ n \end{pmatrix} \qquad (1)$$

Where

$$T = \begin{pmatrix} -2\gamma_1 \omega_1^0 & k_1 \\ -2\gamma_2 \omega_2^0 & k_2 \end{pmatrix} \qquad (2)$$

$\gamma$ is the Grüneisen parameter, $k$ is the doping shift rate and $\omega^o$ is the no-strain and no-doping peak position. The subscript denotes the corresponding Raman modes. In graphene, ω$_1$ subscript *(1)* and ω$_1$*(2)* are G and 2D modes, respectively, where $\gamma_G = 1.95$, $\gamma_{2D} = 3.15$, $k_G = -1.407 \times 10^{-12} cm^{-1}$ and $k_{2D} = -0.285 \times 10^{-12} cm^{-1}$ [5,7]. In fact, the vector space of Raman peak positions $\omega_1$-$\omega_2$ is a linear transformation from the $\varepsilon$-$n$ space, while the origin of both spaces defines the absence of strain and doping. Therefore, ω represents the deviation of the recorded frequency from $\omega^0$ due to strain or doping. It is to be noted that *n* represents the relative shift in the charge carrier and mostly originates from the charge exchange with the substrate. Also, the airborne impurities adsorb over the surface and at edge region may influence *n*.

**S4. Molecular Dynamics and density functional theory:** The atomic scale feature of graphene conformation over textured silicon surfaces is investigated through molecular dynamics (MD) and density functional theory (DFT) calculations of the graphene/Si composite system in different configurations. Initially, we have performed DFT calculations of bulk silicon and monolayer graphene in order to evaluate appropriate MD potential through comparison of structural properties. DFT calculations are performed using the Quantum ESPRESSO software[8,9]. Vanderbilt ultrasoft pseudopotentials[10] were employed, with a wavefunction cutoff of $E_{cut}$ = 80 Ry, a Fermi-Dirac smearing of width 0.01 Ry, and a dense Monkhorst-Pack $k$-point grid sampling of *17x17x17* for bulk Silicon in a cubic supercell and *17x17x1* and for an isolated graphene monolayer in a hexagonal supercell. In the latter case, care is taken to introduce sufficient space between adjacent monolayers by using a large, fixed *z*-direction spacing of 25 Å, so there is no interaction between adjacent periodic images. Our calculations find lattice constants $a_{Si}$ = 0.388 nm and $c_{Si}$ = 0.541 nm in bulk silicon (see **Figure S2**) and $a_{Gr}$ = 0.245 nm for the isolated graphene monolayer, with associated bond lengths $b_{Si-Si}$ = 0.237 nm and $b_{C-C}$ = 0.142 nm.

MD calculations were performed using the LAMMPS MD suite[11]. We have opted for the Stillinger-Weber potential for bulk Silicon[12], parameterized according to the GGA-DFT calculations of Lee and Hwang[13], which provides a better approximation of elastic properties such as the restoring forces on displaced atoms[14,15]. Graphene carbon interactions have been simulated using the LCBOP potential[14]. These potentials are chosen on the basis that they can reproduce DFT and experimental[16,17] lattice constants and bond lengths to a high degree of accuracy, which is crucial in resolving accurate structural properties. For example, our simulations find $b_{Si-Si}$ = 0.238 nm and $b_{C-C}$ = 0.1419 nm for silicon and graphene respectively, which is essentially in perfect agreement to the quantum calculations. The carbon-silicon interaction has been implemented using a Lennard-Jones 6-12 potential with $\varepsilon_{Si-C}$ = 8.909 meV and $\sigma_{Si-C}$ = 0.3326 nm to model physisorption of the graphene monolayer on a silicon substrate[16,18].

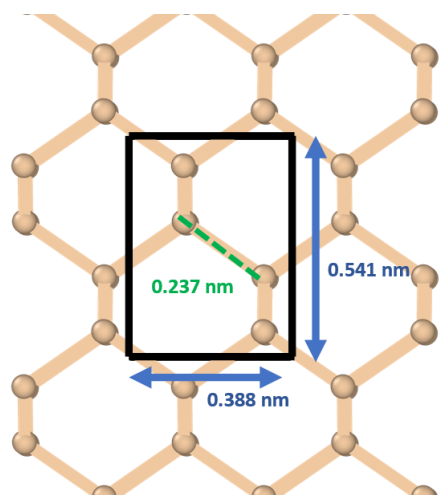

**Figure S2.** DFT-calculated structure of bulk Silicon. Lattice constants and the Si-Si bond length are indicated.

To investigate the effect of different textured silicon surfaces have on adsorption and strain of a graphene monolayer, different approximately Gr/Si-lattice-matched simulation cells are constructed with and without nanometre-scale trenches, introduced into the silicon substrate through the removal of atoms. This necessarily requires the use of large cells in order to obtain satisfactory trench depth and to perform lattice-matching between the graphene and silicon unit cells, along the *x* (long) and *y* (small) axes (as shown in **Figure S3a**). Table 2 lists the graphene and Si cell dimensions and unrelaxed cell lengths for all of the structures we have simulated, as well as the corresponding initial lattice mismatch. For the P40 cell only, we have also considered a number of different lattice-matching conditions along the y-axis in order to check the consistency of our results and find very similar variation in bond and strain properties upon full relaxation. In all cases, the graphene layer is deposited along the (001) surface of silicon, and the graphene sheet is long in the [1010] (armchair) direction. For the largest cells, which are used to model the P250 textured surfaces, the cells employed contain approximately 370,000 atoms. The simulation cell used to model the P40 structure is shown in **Figure S3**.

| Cell | Graphene X | Graphene Y | Silicon X | Silicon Y | $x_{strain}$ (%) | $y_{strain}$ (%) | Depth (nm) |
|---|---|---|---|---|---|---|---|
| **P40** | 39.614 nm [93] | 3.197 nm [13] | 39.670 nm [102] | 3.111 nm [8] | 0.1412 | -1.3773 | 5 |
| **P40** | 39.614 nm [93] | 3.935 nm [16] | 39.670 nm [102] | 3.889 nm [10] | 0.1412 | -0.5869 | 5 |
| **P40** | 39.614 nm [93] | 4.667 nm [19] | 39.670 nm [102] | 4.672 nm [12] | 0.1412 | -0.0600 | 5 |
| **P125** | 125.234 nm [294] | 3.197 nm [13] | 125.234 nm [322] | 3.111 nm [8] | 0.0003 | -1.3773 | 10 |
| **P250** | 250.468 nm [588] | 3.197 nm [13] | 250.468 nm [644] | 3.111 nm [8] | 0.0003 | -1.3773 | 10 |

**Table 2:** Details of different simulation cells used in this work. The cell lengths of the graphene and silicon structures, used for the P40, P125 and P250 textures are shown in their respective columns, with the corresponding number of unit cells shown in square brackets. Equilibrium graphene and silicon lengths in the x- and y-directions, associated initial mismatch, and the depth of the silicon slab are also shown.

In order to approximate the experimental textured surfaces, we have introduced a nanometre-scale trench into the silicon substrate and extrapolated the change in lattice parameters upon geometric optimisation of all atoms for a flat (**Figure S3b**) and textured (**Figure S3c**) silicon slab. A trench of width 20nm and depth 3nm is used for the P40 configuration, while trenches of width 40nm and depth 5nm are used for the P125 and P250 configurations, in order to mimic the experimental configuration.

For all of the flat geometries, we find that the silicon slab causes a compression of the graphene monolayer. Notably, for the flat P125 and P250 geometries, which contain very little strain from lattice-matching of the graphene monolayer to the Si substrate, we find a minimal additional compression of around 0.12% (blue color in z-scale bar) of the average graphene bond length with respect to a perfect free-standing monolayer, which is in very good agreement with the experimental Raman data. Upon relaxation of the suspended geometries, we find configurations which give a close approximation to the partially-conformal morphologies of a graphene monolayer over a grooved substrate[19].

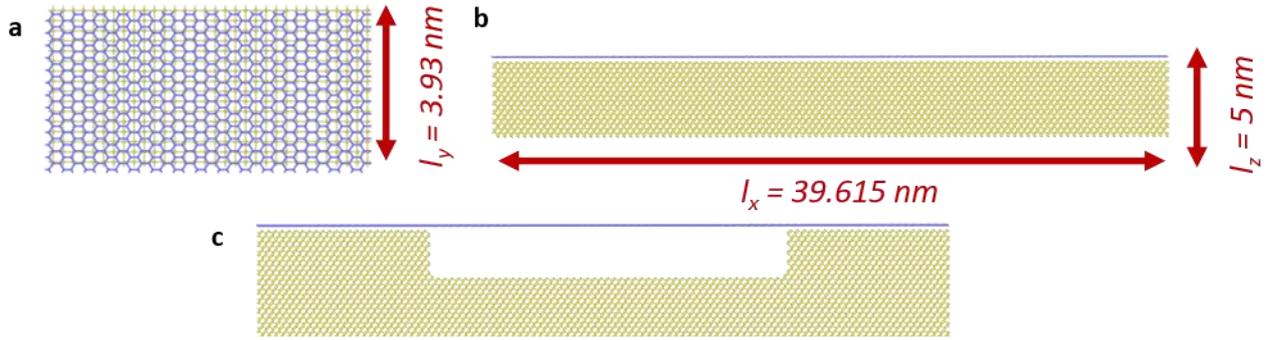

**Figure S3.** a) Top view of the perfect (i.e. without any surface texturing) P40 simulation cell and b) side view. The cell dimensions employed in our simulations are listed, for details of the number of unit cells used in this construction, see Table 2. c) Initial configuration of the P40 graphene/silicon trench structure used to simulate the experimental textured surfaces. The corresponding relaxed geometries are shown in the main text, **Figure. 4c-f**.

We note that there are different distributions of strain across the flat (crest) and buckled (trough) regions, with bond variations. A full analysis of the local variation in the bond and strain distribution for the grooved P40 geometry has been carried out and is shown in the main text **Figure 4c-f**. The bond length has been calculated between adjacent carbon atoms using the formula $b_0 = \sqrt{\Delta x^2 + \Delta y^2 + \Delta z^2}$, and local changes in the parallel and perpendicular strains have been extracted from the local bonds through comparison of the values of the $\Delta x$ and $\Delta y$ bond components to the values of a bond with the same angular orientation with a bond length equal to that of graphene adsorbed on a flat Si surface (i.e. compressed by 0.11% w.r.t free-standing graphene). The net changes in bond length (**Figure 2d** in main text) were calculated by averaging the C-C $b_0$ for carbon atoms in the crest, trough and across the entire cell.

**S5. AFM and FFM calibration:** Calibration of normal and torsional spring constants was done regularly at the beginning and during friction measurements according to Sader method[20,21]. Tip apexes were systematically estimated performing blind tip reconstruction by NT-MDT software (i.e. deconvolution of the topography image obtained on a special calibration grating composed of random distributed nanometric tips, PA01/NM by NT-MDT). Few tips were also imaged by scanning electron microscopy (SEM). Tips height was controlled by SEM images and found equal to the nominal value with a deviation of the order of 5%. The nominal value of 16 nm was used for all our tips.

Blind tip reconstruction and SEM images are presented on **Figure S4**. Tips parameters are presented on Table S2.

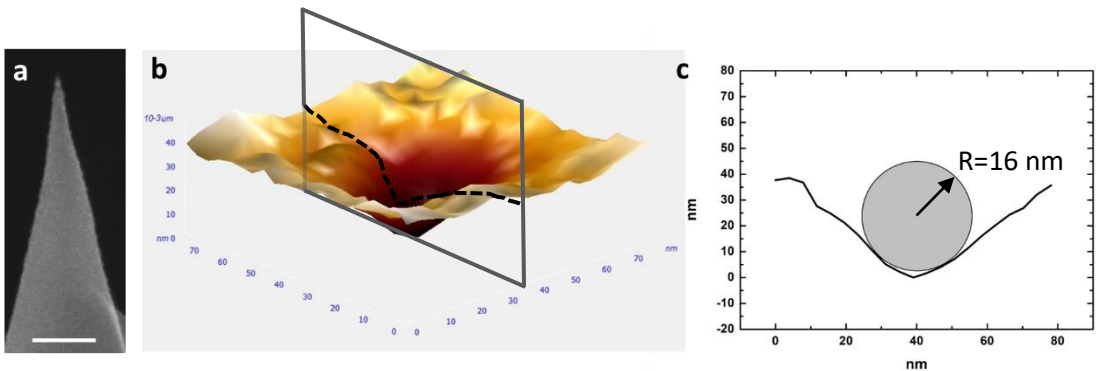

**Figure S4.** a) Representative SEM image of a new tip apex. scale bar 100nm; b) 3D blind tip reconstruction during FFM measurements; c) Tip profile extracted from 3D reconstruction and evaluation of the curvature radius.

**Table S 2: Measured parameters of few relevant tips used during the experiment**

| Tip | $K_{norm}$ (N/m) | $K_{tors}$ (N m) $10^{-8}$ | Radius (nm) | Height (µm) |
|---|---|---|---|---|
| B6_b | 0,26 | 1,06 | 19 ± 1 | 16 ± 1 |
| B7_b | 0,25 | 1,09 | 27 ± 2 | 16 ± 1 |
| B9_b | 0,23 | 1,05 | 21 ± 1 | 16 ± 1 |
| B10_a | 0,76 | 1,53 | 23 ± 1 | 16 ± 1 |
| B11_b | 0,27 | 1,08 | 16 ± 1 | 16 ± 1 |
| B13_b | 0,25 | 1,10 | 19 ± 1 | 16 ± 1 |

Beams Length: a = 250 µm; b = 350 µm

Lateral force refers to a single lateral force signal, forward or backward as acquired from the microscope, properly calibrated[22]. Friction force resulted from the difference between forward and backward lateral signals divided by two (standard TMR analysis). Images or maps were typically acquired at ~1 Hz scan rate, on a 1x1µm² area. To establish the normal load applied, force-distance curves were previously acquired to calculate the photodetector sensitivity along the vertical direction. The same sensitivity was used for calibrating the lateral force signal.

To obtain friction vs load curves, the orthogonal scans were acquired in "one line" mode (512 pts per lines) by decreasing the set point (i.e. decreasing the applied normal load) every forty lines from~30nN to the pull-off value whereas for the parallel scan, an entire image (512x 512 pts) for each normal load was acquired. In the orthogonal scans, the friction forces were averaged on approximately thirty lines corresponding to a constant applied normal load to produce one data point; in the case of parallel scans, friction force values were averaged on a selected area of interest containing a number of points comparable to that used for orthogonal friction data points.

All the areas analyzed have been previously mechanically cleaned with a cantilever different from that used for friction measurement, using a normal load (~50nN) greater than the highest load reported in the plot.

AFM image processing, including the three-dimensional display of data, was carried out using both the software provided by NT-MDT and the free modular software Gwyddion (version 2.55).

The typical friction responses of graphene deposited on flat region investigated in two different orientations are reported in **Figure S5**. 3D topography reconstruction (**Figure S5a**) highlights the presence of sub-nanometeric corrugations over the graphene surface; lateral force map (S5b) as well as its profile (S5c) show an almost constant trend without the occurrence of any periodic modulation. A very similar behavior has been observed on the same region rotated by 90 degrees (**Figure 3 d,e,f**), confirming thus the complete isotropic behavior of the system.

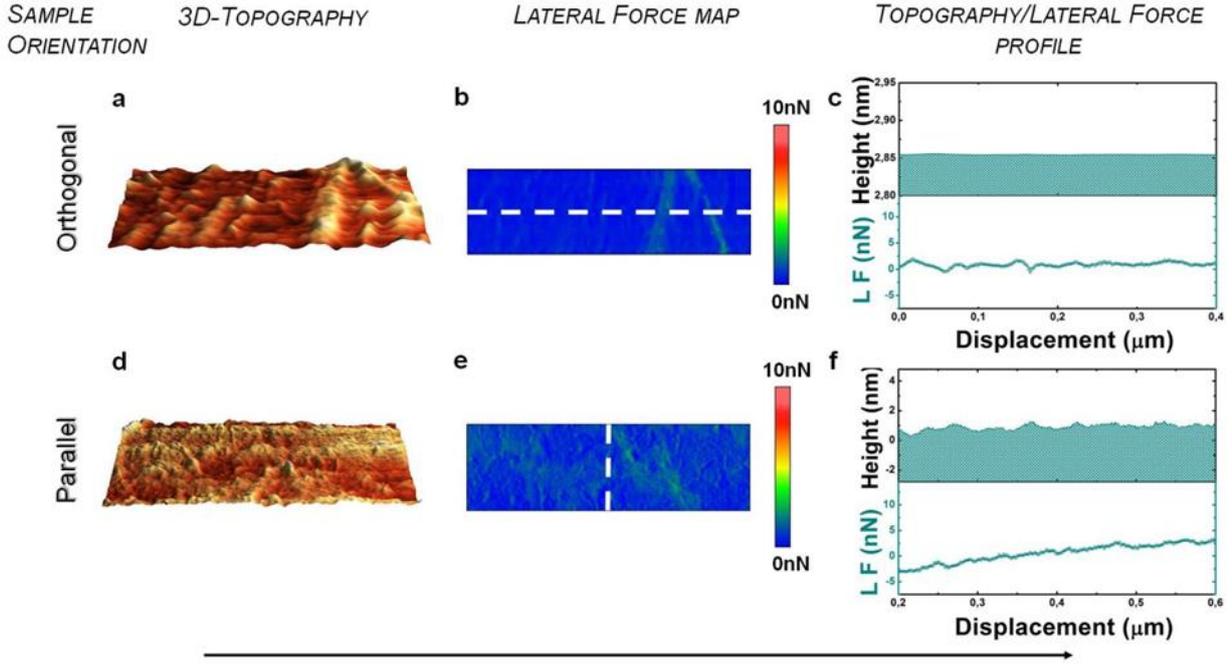

**Figure S5. Effect of scan direction on lateral force for Gr flat region.** (a) 3D topography image (1.0 x 0.5 micron) and (b) lateral force map (1.0 x 0.3 micron) measured simultaneously on flat region graphene-covered. Applied load ≈25nN. (c) Top, height profile and, bottom, corresponding lateral force (open dots correspond to raw data, dark cyan line is the result of 7 pts smoothing) extracted from white dashed line in (b); length 400nm. (d) 3D topography image (1.0 x 0.5 micron) and (e) lateral force map (1.0 x 0.3 micron) measured simultaneously on flat region graphene-covered rotated by 90° degrees. Applied load ≈25nN. (f) Top, height profile and, bottom, corresponding lateral force profile extracted from white dashed line in (e); length 400nm.

**S6. Friction vs load curves fitting procedure:** The procedure utilized to fit friction versus load curves was tested on different flat regions next to the textured areas. The measured friction force $F_f$ is displayed according to method developed and described by Carpick *et al.*[23] where the square root of $F_f$ normalized to $F_o$ (friction force measured at zero applied load) is plotted as a function of external applied load $L_{ext}$

$$\sqrt{\frac{F_f}{F_0}} = \left(\frac{a + \sqrt{1 - \frac{L_{ext}}{L_0}}}{1 + a}\right)^{2/3}$$

The square root is because formula refers originally to contact radius variation with load (sphere on a flat surface) but using single asperity approximation ($F_f = \tau \cdot \pi r^2$) the formula can be rewrite in terms of friction forces which are our measurable. An example is presented on **Figure S6**. Data are fitted leaving *a* and $L_o$ (the pull-off force in a force-distance cycle) as a free parameters. The significance of *a* is to evaluate the contact behavior with respect to the JKR model (Johnson-Kendall-Roberts) corresponding to *a* close to one and to DMT model (Derjaguin-Mueller-Toporov) corresponding *a* close to zero. Finally, $L_0$ can be compared to experimental measurements and results from DMT data fitting in order to test consistency of the procedure. Test performed on 5 different regions always reveals a < 0.05 confirming the use of DMT approximation.

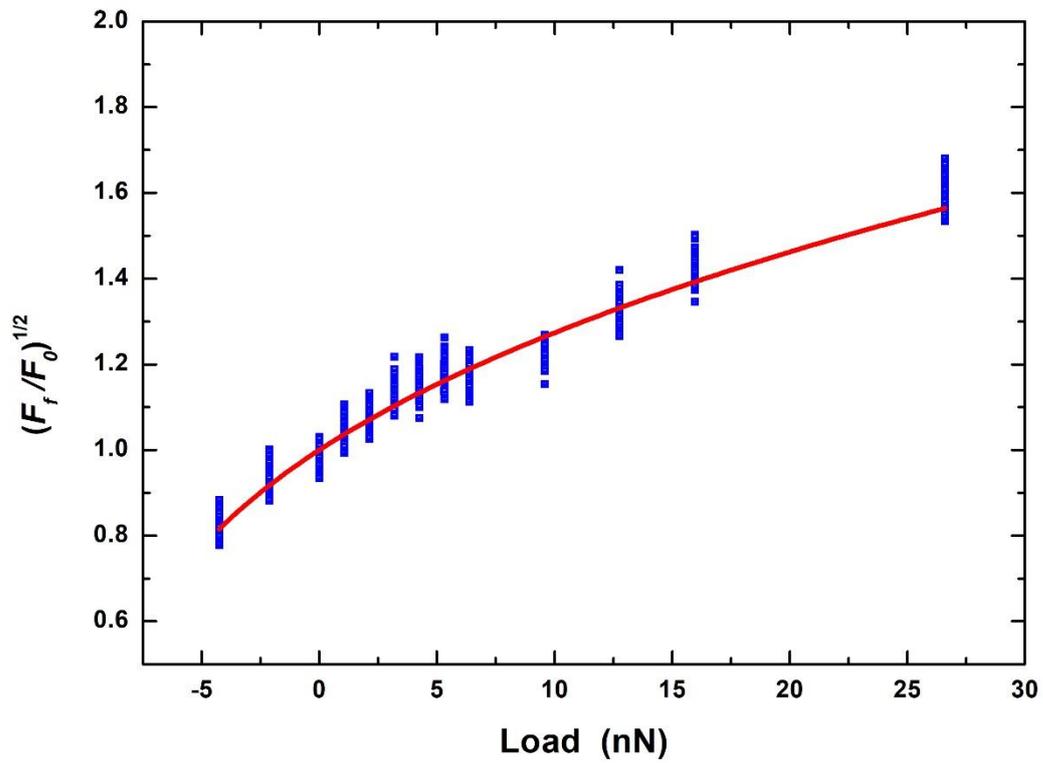

**Figure S6.** Blue dots represent the experimental data; Red curves is the fit according to method developed and described by Carpick et al.[23] where the square root of $F_f$ normalized to $F_o$ (friction force measured at zero applied load) is plotted as a function of external applied load $L_{ext}$. In this case we have $F_o = 4.25*10^{-10}$ N and we obtain as fitting results $a = 0.001$ and $L_o = (-9.2 +- 0.7)*10^{-9}$ N ($L_o$ is the pull-off force in a force-distance cycle).

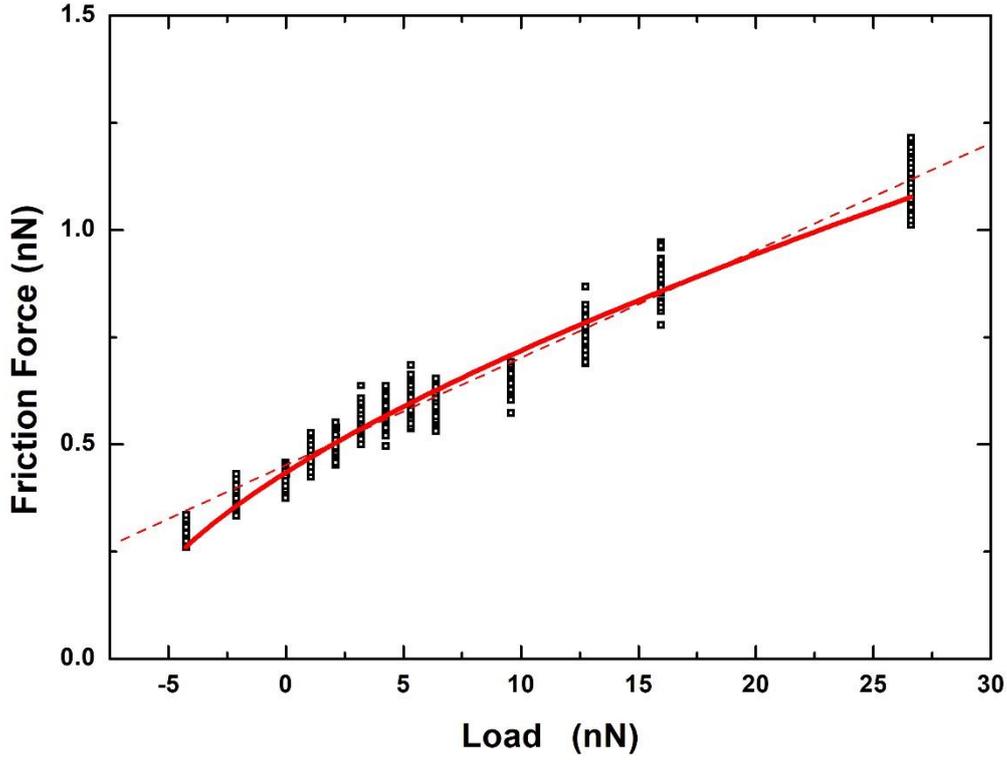

**Figure S7.** Black dots represent the raw data; red curve is the fitting curve according to DMT model + offset, red dashed line is the linear fitting.

In **Figure S7** raw data corresponding to curve at **Figure S6** and the corresponding DMT fitting curve are presented.

$$F_f = y_0 + A \cdot (L_{ext} + L_0)^{2/3}$$

where $L_{ext}$ represent the external load, $L_0$ is the adhesion force and we add the offset $y_0$ to accounts for small but unavoidable non zero values of the photodetector lateral signal at zero applied load (i.e. far from the surface).

During fitting procedure the power law exponent is fixed at 2/3 and eventually $L_0$ was forced to the experimental pull off force value obtained by force – distance curve so that the outcome of this procedure is only the A parameter ($A = S \cdot \pi \cdot \left(\frac{3}{4} \cdot \frac{R}{K}\right)^{2/3}$) that contains information about shear strength S at the interface.

To evaluate quantitatively the shear strength S the reduced Young's modulus of the interface (K) and the tip radius R have to be estimated. We used SEM images of tips apex and we performed blind tip reconstruction in connection with all friction vs load curves to evaluate R (see tab. 1 above). On the contrary the reduced Young's modulus K is estimated from tabulated values of mechanical parameters of the materials involved. The Young's modulus (E = 70 GPa) and Poisson's ratio (ν = 0.2) of the $SiO_2$ has been used for the tip apex and bare substrate[24]. Instead we used, in analogy with literature results[24,25] the bulk elastic constants of graphite (E = 30 GPa; ν = 0.24), for the single layer graphene deposited on $SiO_2$. Following these approximations, we extract a shear strength of 25 MPa from the curve S6. Similarly, we obtain the results summarised on Table 2 in the main text and the complimentary analysis on GrP(40) sample presented on **Figure S8**.

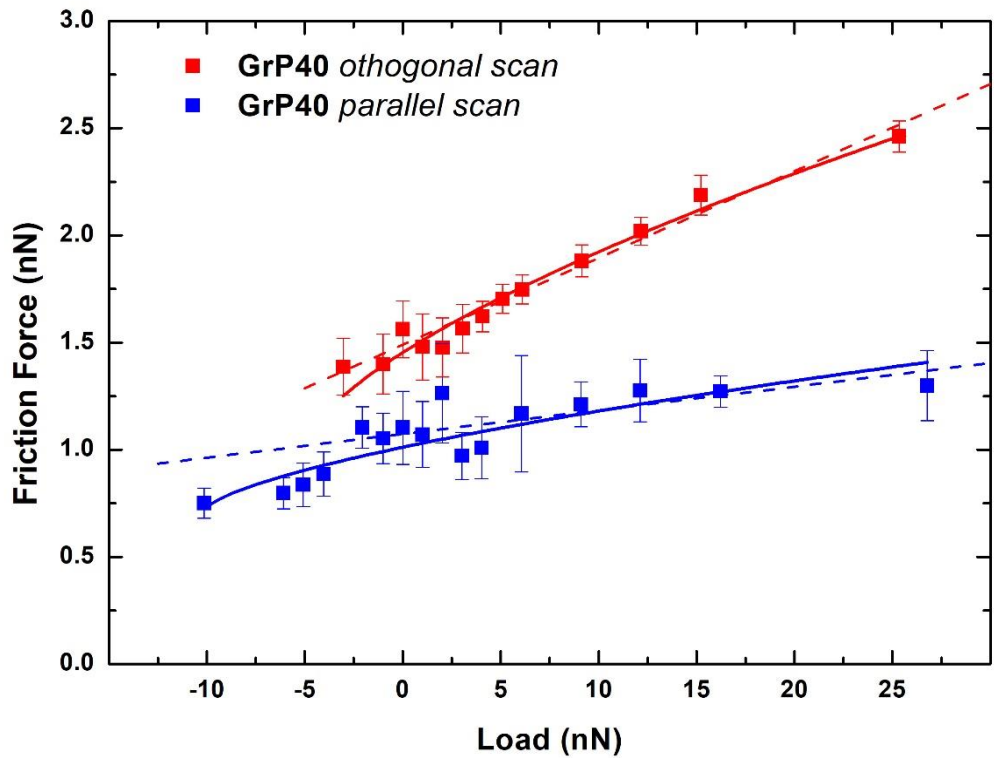

**Figure S8.** Load depended friction force on GrP40 sample with the grooves axis oriented parallel (blue) and perpendicular (red) to the fast scan direction. Analyzed area corresponds to a completely different region with respect to the one analyzed in Figure 4 in the main text. Square dots represent the raw data while continuous lines are the fitting curve according to DMT model. The corresponding shear strength are 43 MPa and 17 MPa for sliding direction perpendicular and parallel to the texture respectively

**References**

(1)   Dell'Anna, R.; Iacob, E.; Tripathi, M.; Dalton, A.; Böttger, R.; Pepponi, G. *J. Microsc.* **2020**, No. 1, jmi.12908.

(2)   Dell'Anna, R.; Masciullo, C.; Iacob, E.; Barozzi, M.; Giubertoni, D.; Böttger, R.; Cecchini, M.; Pepponi, G. *RSC Adv.* **2017**, *7* (15), 9024–9030.

(3)   Sabio, J.; Seoánez, C.; Fratini, S.; Guinea, F.; Neto, A. H. C.; Sols, F. *Phys. Rev. B - Condens. Matter Mater. Phys.* **2008**, *77* (19), 1–8.

(4)   Aitken, Z. H.; Huang, R. *J. Appl. Phys.* **2010**, *107* (12), 123531.

(5)   Lee, J. E.; Ahn, G.; Shim, J.; Lee, Y. S.; Ryu, S. *Nat. Commun.* **2012**, *3*, 1024.

(6)   Michail, A.; Delikoukos, N.; Parthenios, J.; Galiotis, C.; Papagelis, K. *Appl. Phys. Lett.* **2016**, *108* (17), 173102.

(7)   Jiang, T.; Wang, Z.; Ruan, X.; Zhu, Y. *2D Mater.* **2018**, *6* (1), 015026.

(8) Giannozzi, P.; Baroni, S.; Bonini, N.; Calandra, M.; Car, R.; Cavazzoni, C.; Ceresoli, D.; Chiarotti, G. L.; Cococcioni, M.; Dabo, I.; Dal Corso, A.; De Gironcoli, S.; Fabris, S.; Fratesi, G.; Gebauer, R.; Gerstmann, U.; Gougoussis, C.; Kokalj, A.; Lazzeri, M.; Martin-Samos, L.; Marzari, N.; Mauri, F.; Mazzarello, R.; Paolini, S.; Pasquarello, A.; Paulatto, L.; Sbraccia, C.; Scandolo, S.; Sclauzero, G.; Seitsonen, A. P.; Smogunov, A.; Umari, P.; Wentzcovitch, R. M. *J. Phys. Condens. Matter* **2009**, *21* (39).

(9) Giannozzi, P.; Andreussi, O.; Brumme, T.; Bunau, O.; Buongiorno Nardelli, M.; Calandra, M.; Car, R.; Cavazzoni, C.; Ceresoli, D.; Cococcioni, M.; Colonna, N.; Carnimeo, I.; Dal Corso, A.; de Gironcoli, S.; Delugas, P.; DiStasio, R. A.; Ferretti, A.; Floris, A.; Fratesi, G.; Fugallo, G.; Gebauer, R.; Gerstmann, U.; Giustino, F.; Gorni, T.; Jia, J.; Kawamura, M.; Ko, H.-Y.; Kokalj, A.; Küçükbenli, E.; Lazzeri, M.; Marsili, M.; Marzari, N.; Mauri, F.; Nguyen, N. L.; Nguyen, H.-V.; Otero-de-la-Roza, A.; Paulatto, L.; Poncé, S.; Rocca, D.; Sabatini, R.; Santra, B.; Schlipf, M.; Seitsonen, A. P.; Smogunov, A.; Timrov, I.; Thonhauser, T.; Umari, P.; Vast, N.; Wu, X.; Baroni, S. *J. Phys. Condens. Matter* **2017**, *29* (46), 465901.

(10) Giannozzi, P.; Baseggio, O.; Bonfà, P.; Brunato, D.; Car, R.; Carnimeo, I.; Cavazzoni, C.; De Gironcoli, S.; Delugas, P.; Ferrari Ruffino, F.; Ferretti, A.; Marzari, N.; Timrov, I.; Urru, A.; Baroni, S. *J. Chem. Phys.* **2020**, *152* (15).

(11) Garrity, K. F.; Bennett, J. W.; Rabe, K. M.; Vanderbilt, D. *Comput. Mater. Sci.* **2014**, *81*, 446–452.

(12) Plimpton, S. *J. Comput. Phys.* **1995**, *117* (1), 1–19.

(13) Stillinger, F. H.; Weber, T. a. *Phys. Rev. B* **1985**, *31* (8), 5262–5271.

(14) Lee, Y.; Hwang, G. S. *Phys. Rev. B* **2012**, *85* (12), 125204.

(15) Jing, Y.; Hu, M.; Guo, L. *J. Appl. Phys.* **2013**, *114* (15), 153518.

(16) Los, J. H.; Fasolino, A. *Phys. Rev. B* **2003**, *68* (2), 024107.

(17) Graziano, G.; Klimeš, J.; Fernandez-Alonso, F.; Michaelides, A. *J. Phys. Condens. Matter* **2012**, *24* (42), 424216.

(18) Kaftory, M.; Kapon, M.; Botoshansky, M. In *Wiley, Chichester*; 1998; Vol. 121, pp 181–265.

(19) Ong, Z.-Y.; Pop, E.; Shiomi, J. *Phys. Rev. B* **2011**, *84* (16), 165418.

(20) Sader, J. E.; Chon, J. W. M.; Mulvaney, P. *Rev. Sci. Instrum.* **1999**, *70* (10), 3967–3969.

(21) Green, C. P.; Lioe, H.; Cleveland, J. P.; Proksch, R.; Mulvaney, P.; Sader, J. E. *Rev. Sci. Instrum.* **2004**, *75* (6), 1988–1996.

(22) Gnecco, E.; Pawlak, R.; Kisiel, M.; Glatzel, T.; Meyer, E. In *Nanotribology and Nanomechanics*; Springer International Publishing: Cham, 2017; pp 519–548.

(23) Carpick, R. W.; Ogletree, D. F.; Salmeron, M. *Area* **1999**, *400*, 395–400.

(24) Buzio, R.; Gerbi, A.; Uttiya, S.; Bernini, C.; Del Rio Castillo, A. E.; Palazon, F.; Siri, A. S.; Pellegrini, V.; Pellegrino, L.; Bonaccorso, F. *Nanoscale* **2017**, *9* (22), 7612–7624.

(25) Deng, Z.; Klimov, N. N.; Solares, S. D.; Li, T.; Xu, H.; Cannara, R. J. *Langmuir* **2013**, *29* (1), 235–243.